\def\gte{\lower 0.5ex\hbox{${}\buildrel>\over\sim{}$}}
\def\lte{\lower 0.5ex\hbox{${}\buildrel<\over\sim{}$}}

\documentstyle[12pt,aasms4]{article}

\begin{document}

\title{Electron Acceleration and the Production of Nonthermal
Electron Distributions in Accretion Disk Coronae}

\author{Hui Li\altaffilmark{1} and James A. Miller\altaffilmark{2}}

\altaffiltext{1}{NIS-2, MS D436, Los Alamos National Laboratory, Los Alamos,
NM 87545; hli@lanl.gov}
\altaffiltext{2}{Department of Physics, The University of Alabama in
Huntsville, Huntsville, AL 35899; miller@mpingo.uah.edu}

\begin{abstract}

We consider electron acceleration by obliquely propagating fast mode
waves in magnetically dominated accretion disk coronae. For low
coronal plasma densities, acceleration can exceed Coulomb drag at
lower energies and energize electrons out of the thermal background,
resulting in a nonthermal tail. The extent of this tail is determined
by the balance between acceleration and radiative cooling via inverse
Compton scattering and synchrotron emission, and usually goes out to 
tens of MeV. This will have direct applications for explaining the
gamma-rays from several galactic black hole candidates, such as Cyg
X-1 and GRO J0422, which show $0.5$--$5$ MeV emissions in excess over
what most thermal models predict.  Detailed time evolutions of the
particle distributions and wave spectra are also presented. 

\end{abstract}


\section{Introduction}

Thermal Comptonization models have had much success in explaining the
hard X-ray spectra from galactic black hole candidates (GBHCs) with
the plasma temperature $\sim 50$-$100$ keV and Thomson depths of a
few (e.g., \cite{sle76}; \cite{st80}; \cite{har94}; \cite{lia93}). 
However, the most sensitive observations of GBHCs to date in the 
$0.5$--$5$ MeV range by {\it Compton} Gamma-Ray Observatory have
clearly revealed that persistent gamma rays ($>1$ MeV) are being
produced in some GBHCs, notably Cyg X-1 (\cite{mcc96}; \cite{ph96};
\cite{ling96}) and GRO J0422 (\cite{dij95}). These gamma-ray
emissions are very difficult to accommodate by the pure thermal
models (e.g., \cite{st80}; \cite{t94}), strongly suggesting the need
for modification (\cite{sd95}) or to incorporate some nonthermal
processes.

Recently, Li, Kusunose, \& Liang (1996) have proposed a gyroresonant
stochastic electron acceleration model to account for the MeV 
emissions from GBHCs (see also Dermer, Miller, \& Li 1996). In that
model, they showed that high frequency whistlers can accelerate
electrons directly from the thermal background, after which Alfv\'en
waves would continue the acceleration to higher energies. The result
is an electron distribution with a hard non-Maxwellian tail. Compton
scatterings from both thermal and nonthermal electrons produce a
broad band X- to gamma-ray spectrum, in agreement with that observed
from both Cyg X-1 and GRO J0422.  One uncertainty associated with
that model is the generation of whistlers, which can arise from
either a cascade of wave energy from lower frequencies or a
microinstability (see e.g., \cite{gary93}). While the nature of wave
cascading is fairly well known in the MHD regime (e.g., \cite{v94};
\cite{rw95}), it has not been investigated at higher frequencies.
Producing waves that can gyroresonate with an electron of a given
energy through a resonant microinstability requires (among other
things) an anisotropic electron distribution containing electrons of
that energy. Nonresonant instabilities may also produce the needed
waves, but still require specific anisotropic distributions.

In this {\it Letter}, we consider electron acceleration by MHD fast
mode waves that likely exist in accretion disk coronae. 
These low-frequency waves can accelerate electrons from
thermal to relativistic energies. Coupled 
time-dependent diffusion equations for the electron distribution and
the wave spectral density are solved numerically (\cite{mlm96},
hereafter MLM). Nonthermal electron distributions are clearly
obtained under a range of parameters in the accretion disk
environment, which will have a direct bearing on the future 
modeling of the hard X-ray and gamma-ray emissions from GBHCs.

\section{THE MODEL}

\subsection{Basic Model Assumptions and Parameters}

The hydrogen plasma flowing around a black hole with mass $M$ is 
assumed to form an accretion disk (e.g., \cite{ss73}), from which
most of the soft photons (a few keV) originate. In the inner part of
the disk, we assume that the magnetic energy density $U_{\rm B} =
B_0^2/8\pi$ is in equipartition with the plasma thermal pressure $n_0
k_B T_p$, where $k_{\rm B}$ is Boltzmann's constant; $T_p = T_i +
T_e$; and $T_i$ and $T_e$ ($\sim 50$ keV) are the proton and electron
temperatures, respectively. $T_i$ is, unfortunately, poorly
determined and is chosen to be 10 MeV in this study. The plasma
density $n_0$ is taken to be $\sim 1/(\sigma_T R)$, where
$\sigma_{T}$ is the Thomson cross section.  Since $T_i \gg T_e$, it
roughly corresponds to the two-temperature accretion disk situation. 
However, there have been discussions in the literature questioning
whether a two-temperature plasma can occur at all in the accretion
plasma (\cite{phin81}; \cite{rees82}; \cite{bc88}). We emphasize that
we use $T_i = 10$ MeV only to get a fiducial magnetic field value,
and in fact, our model works even better for an isothermal accretion
disk because electrons have a much higher thermal speed than protons
and are preferentially accelerated (see below). 

A tenuous, quasi-spherical extended corona surrounding the hot inner
disk is postulated (e.g., \cite{lp77}; \cite{hm91}; \cite{hmg94}). 
The coronal electron temperature is again taken to be 50 keV initially,
and the coronal optical depth $\tau_c = n_c \sigma_T R \lte 1$ is varied, 
where $n_c$ is the coronal electron density. We further assume 
that the magnetic field in the corona is the same as that
in the disk $B_0$, so that the corona is magnetically dominated and
the plasma $\beta \equiv n_c k_B T_p / U_{\rm B}$  is $\leq 1$.
Notice that $\beta = \tau_c$ with these conditions.
Adopting the black hole mass $M$ to be $10M_{\odot}$ and the size $R$
of the system to be $\approx 30 GM/c^2$, within which most of high
energy emissions are produced, we find that the coronal magnetic 
field $B_0$ and dimensionless Alfv\'en speed $v_{\rm A}/c$ are $\sim
3.7 \times 10^6$ G and $B_0 /\sqrt{4\pi n_c m_p c^2} = 0.15\beta^{-1/2}$,
respectively. The $\beta < 1$ condition also
implies that the proton thermal speed is always less than $v_{\rm A}$.

Here, we make a key assumption that a fraction of the total available
energy goes into generating MHD turbulence. This wave turbulence
consists of fast mode and shear Alfv\'en waves (the slow mode will be
heavily Landau damped and will probably not be excited), but it is
just the fast mode waves that are relevant for electron acceleration.
Fast mode waves propagating obliquely with respect to the ambient
magnetic field have a parallel magnetic field component, which can
couple strongly (or resonate) with a particle when the parallel phase
speed of the wave $\omega/k_\parallel$ is about equal to the parallel
component of particle velocity $v_\parallel$; i.e., when $v_\parallel
= v_{\rm A}/\eta$, where we have used the dispersion relation for the
fast mode waves and $\eta$ is the cosine of the wave propagation
angle. This process is referred to as transit-time damping or
magnetic Landau damping (e.g., \cite{s92}; \cite{lv75};
\cite{ach81}), and a broad-band wave spectrum leads to both rapid
wave damping and particle acceleration (MLM). This process is
essentially the resonant form of Fermi acceleration (\cite{fermi49}).
Particle interactions with the parallel magnetic field variations
(caused by the waves) can be viewed as either head-on (gaining
energy) or trailing (losing energy) ``collisions''. Head-on
collisions occur more often than trailing ones, so that net
acceleration results. The resonant condition implies that the
acceleration threshold is $v_{\rm A}$. For the coronal plasma,
$v_{\rm A}$ is comparable to the electron thermal speed but greater
than the proton thermal speed if $\beta$ is less than 1, so that
electrons are preferentially accelerated.

\subsection{The Electron and Wave Diffusion Equations and 
Energy Transfer Rates}

To simplify the calculations, we consider an isotropic fully-ionized
hydrogen plasma permeated by a homogeneous background magnetic field
$B_0$. At some large scale $\lambda_{\rm inj}$, an unspecified
mechanism generates fast mode waves.  The evolution of the
electron distribution $N(E)$ is given by the
Fokker-Planck equation
\begin{equation}
\label{eq-par-dis}
{\partial N \over \partial t} = -{\partial \over \partial E}\left\{
\left[\Big\langle{dE \over dt}\Big\rangle_{\rm acc}
+ \left({dE \over dt}\right)_{\rm loss} \right] N\right\} +
{1 \over 2}{\partial^2 \over \partial E^2}
\left[(D + D_c) N \right] \quad , 
\end{equation}

\noindent where $E$ is the kinetic energy. Here, we have neglected 
the escape and (possible) $e^{-}$--$e^{+}$ pair production.  Those 
coefficients associated with wave-particle interactions are the 
systematic mean acceleration rate $\langle dE/dt \rangle_{\rm acc} =
p^{-2}\partial [p^2 v D(p)]/\partial p$ and the diffusion coefficient
$D = 2 v^2 D(p)$, where $v$ and $p$ are the electron speed and
momentum, and $D(p)$ is the momentum diffusion coefficient (given
below). The other convection term $(dE/dt)_{\rm loss}$ represents the
sum of electron energy change rates from inverse Compton scattering and 
synchrotron (ICS) losses $(dE/dt)_{\rm ics}$ and $e$-$e$ Coulomb collisions
$(dE/dt)_{\rm c}$ (\cite{dl89}), which also give rise to diffusion 
$D_{\rm c}$ (\cite{spi62}; \cite{dl89}).  These are the most important 
processes for electrons in our parameter regime. The processes
that are neglected include: the electron-proton Coulomb interaction
since it is much slower than $e$-$e$ (when $T_i < 100$ MeV);  
the diffusion due to Compton scatterings since the electron recoil is
typically small and wave-particle diffusion will dominate at high
energies; the energy gain due to the convergence of the flow since 
the accretion time is much longer than dynamic timescale ($\sim R/c$).  
The momentum diffusion
coefficient $D(p)$ for transit-time damping is (MLM)
\begin{equation}
\label{eq-dpp}
D(p)~=~(m_ec)^2~{\pi \over 16}~\left({v_{\rm A}\over c}\right)^2~
c\langle k \rangle~\zeta~\left({p\over m_ec}\right)^2
\left({c\over v}\right)~F(\mu_0)
\end{equation}

\noindent where $m_e$ is the electron mass, $\zeta = U_{\rm T} /
U_{\rm B}$, $U_{\rm T}$ is the fast mode wave energy density,
$\langle k \rangle$ is the mean wavenumber of the fast mode waves, 
$\mu_0 = v_{\rm A} /v$ and $F(\mu_0)$ is basically an efficiency
factor, which equals zero when $v \le v_{\rm A}$ (i.e., when
resonance is impossible) and equals $- 5/4 - (1+2\mu_0^2)\ln\mu_0 +
\mu_0^2 + {1\over 4} \mu_0^4$ otherwise. 

The evolution of the isotropic wave spectral density $W_{\rm T}$ can
be approximated by (see \cite{zm90})
\begin{equation}
\label{eq-wave-dis}
{\partial W_{\rm T} \over \partial t} = {\partial \over \partial k}\left[
k^2 D_{\rm W} {\partial \over \partial k}\left( k^{-2} W_{\rm T} \right) 
\right] -\gamma W_{\rm T} + Q_{\rm W}\delta(k-k_0) \quad ,
\end{equation}

\noindent where we have included a term $\gamma (k)$ 
for the wave damping 
by the electrons and a term $Q_{\rm W}$ for the injection. 
The wave damping rate $\gamma (k)$ can be obtained from the relation
$\int_0^{\infty}dk \gamma(k) W_{\rm T}(k) = \int_0^{\infty}dE
N(E) \langle dE/dt \rangle_{\rm acc}$ (cf. equation (\ref{eq-dpp})).
Since $\langle dE/dt\rangle_{\rm acc} \propto \langle k \rangle$,
this implies that large scale waves (small $k$) will suffer little loss, 
and that rapid wave dissipation only occurs when $k$ is sufficiently large. 

At steady state, the volumetric wave energy injection rate $Q_{\rm
W}$ [ergs cm$^{-3}$ s$^{-1}$] at $k_0 = 2\pi/\lambda_{\rm inj}$ must
equal to the rate at which energy is transferred to smaller scales
$\sqrt{2} v_A k_0 \zeta^{3/2} U_{\rm B}$ (MLM). This energy flow is
eventually dissipated at higher wavenumbers by electrons, which in
turn produce X-ray to gamma-ray emissions. This implies that the
volumetric gamma-ray energy production rate $Q_{\gamma} \sim
L_{\gamma} / (4\pi/3R^3)$, where $L_{\gamma}$ is the gamma-ray
luminosity, must be smaller or equal (steady state) to $Q_{\rm W}$.
This gives
\begin{equation}
\label{eq-ugub}
{U_{\gamma} \over U_{\rm B}} \lte 3~{v_A\over c}
{R \over \lambda_{\rm inj}}~ \zeta^{3/2} \quad ,
\end{equation}

\noindent where $U_{\gamma} = L_{\gamma} / (4\pi R^2 c)$ is the
energy density of gamma-ray photons.  Letting $\lambda_{\rm inj}/ R
\sim 0.1$, we obtain $U_{\gamma} / U_{\rm B} \lte 4.5 \beta^{-1/2} 
\zeta^{3/2}$. This implies that in order to get $U_{\gamma} / U_{\rm B}
\sim 0.1$ as suggested by the $60$--$1000$ keV luminosity from 
Cyg X-1 (\cite{ph96}), $\zeta$ must be $\approx 0.08 \beta^{1/3}$, 
well within the weak turbulence limit.  

Next, we look at the electron energy gain and loss rates.  The mean
acceleration rate of electrons by the waves is given as (MLM)
\begin{equation}
\label{eq-dedt-acc}
\langle{d\gamma \over dt}\rangle = 
{\pi \over 4}~\left({v_A\over c}\right)^2~
c\langle k \rangle~\zeta\left({p\over m_e c}\right)~G(\mu_0)
\end{equation}

\noindent where $G(\mu_0) = F(\mu_0)+ (4\mu_0^2\ln\mu_0 - \mu_0^4 +
1) /(4\gamma^2)$ when $\mu_0 < 1$ and equals 0 otherwise.  The
product $\langle k \rangle \zeta$ has to be obtained from the
simulations (see next section).  The ICS losses can be written as
$(d\gamma / dt)_{\rm ics} = - (4/3) (\Theta_p / \tau_{\rm dyn}) (1 +
U_{\rm ph}/U_{\rm B}) (p / m_e c)^2$, where $\tau_{\rm dyn} = R/c$
and $\Theta_p = k_B T_p / m_e c^2$. The soft photon energy density
$U_{\rm ph}$ is fixed to be the same as $U_{\rm B}$ in this study.
Note that this loss rate may be an overestimate for mildly 
relativistic particles, thus more careful treatment of the losses
will help the acceleration.

When the Coulomb loss timescale is longer than both the
acceleration and the ICS cooling
timescales, we can define a critical energy $\gamma_c$ at which
acceleration balances radiative cooling, 
$\langle d\gamma/dt \rangle = |(d\gamma/dt)_{\rm ics}|$, 
which gives
\begin{equation}
\label{eq-gc}
\gamma_c \approx 1.3 \times 10^4 ({R\over 4.5 \times 10^7})
\langle k \rangle \zeta \quad ,
\end{equation}

\noindent where we have utilized the fact that
when $p/(m_ec) \sim \gamma_c \gg 1$, $G(\mu_0)/\tau_c$ is almost constant
(ranging from $0.76$--$1.1$) for $0.1 \leq \tau_c \leq 1$.
In order to get substantial acceleration---say
$\gamma_c \sim 10$---it must be that $\langle k \rangle~\zeta$ 
is $> 8 \times 10^{-4}$ (cm$^{-1}$). This offers a direct test of our
simulations. 

\section{Results}
\label{sec-resul}

We solve equations (\ref{eq-par-dis}) and (\ref{eq-wave-dis}) using
the Crank-Nicholson method (MLM), and concentrate on the time
evolution of particles and waves from the start of wave injection
until the steady state is reached, during which waves are constantly
injected at $k_0 = 2\pi / (0.1 R)$. This period turns out to be
always less than or comparable to the dynamic timescale $\tau_{dyn}
\sim 1.5\times 10^{-3}$ sec.  This validates our assumption of
neglecting escape. The Coulomb loss is fixed at the rate that is
corresponding to the initial Maxwellian. Even though this treatment
is not self-consistent, our particle acceleration results should not
be affected much since Coulomb loss plays a negligible role compared
to the ICS cooling at relativistic energies.  The Kolmogorov
phenomenology for the wave evolution is assumed so that $W_{\rm T} =
W_0 k^{-5/3}$ with $W_0 \propto (Q_{\rm W}/v_A)^{2/3} U_{\rm
B}^{1/3}$.  The injection rate $Q_{\rm W}$ is chosen as $\approx
2.8\times 10^{15} (v_A/c)$ ergs cm$^{-3}$ s$^{-1}$, corresponding to 
$\zeta = 0.2$ {\it at steady state}, and $U_{\rm T} = \int_0^{\infty}
W_0 k^{-5/3} dk$ should be a constant for all the cases considered
here; there are all confirmed by the simulations to within the
numerical error. 

Figure 1 summarizes the time evolution of the particle density
distribution $N(E)$ (upper panels) as a function of kinetic energy 
$E$, and the corresponding wave spectral density $W_{\rm T}$ (lower
panels) as a function of wavenumber $k$.  Three different densities
are considered. Each plot has 16 curves in it, corresponding to 5
evenly-spaced time intervals in each of three periods, $t=0$--$0.04
\tau_{dyn}$, $0.04$--$0.16 \tau_{dyn}$, and $0.16$--$0.6 \tau_{dyn}$,
respectively. The particle distributions soften first (e.g., curves
$1-8$ in $\tau_c = 0.5$ case), due to the fact that acceleration is
very inefficient initially since waves have not fully cascaded (i.e.,
small $\langle k \rangle$ as shown in lower panel), and Coulomb and
ICS losses dominate at high energies. As waves cascade over the
inertial range, $\langle k \rangle$ quickly grows to a level that
acceleration overcomes both Coulomb and ICS losses.  Electrons are
then energized out of the thermal background and the nonthermal hard
tail forms. This is indicated, for example, by curves $6$--$15$ in the
$\tau_c = 0.1$ case.  After that, both the particle and wave spectra
gradually reach steady states. 

The $\tau_c = 0.1$ case clearly illustrates several other points.
From simulations, $\langle k \rangle \zeta \sim 10^{-3}$ (cm$^{-1}$),
thus $\gamma_c \sim 10$ using equation (\ref{eq-gc}), which is in
perfect agreement with curve 15. Furthermore, the nonthermal tails
start to develop only at $E/m_e c^2 \sim 0.13$ (corresponding to
$v_{\rm A}/c = 0.46$), complying with the acceleration threshold.
The threshold energies for $\tau_c = 1$ and $0.5$ are buried in the
thermal distributions.

That the particle's high energy cutoff gets larger as $\tau_c$
decreases can be understood from equation ({\ref{eq-gc}). As $\tau_c$
decreases, fewer particles are available to absorb the wave energy,
which results in larger $\langle k \rangle$.  Smaller $\tau_c$ also
reduces the number of particles with $v > v_{\rm A}$, as evident from
the upper panels in Figure 1.

Figure 2 shows the fraction of electrons with $E \geq 511$ keV out of
the total electron population as a function of coronal optical depth
$\tau_c = 0.1$--$1$, both in energy content $f_{E511} =
\left[\int_{511}^{\infty}E N(E)dE \right] / \left[\int_{0}^{\infty}E
N(E)dE \right]$ (upper panel), and in number $f_{N511} = 
\left[\int_{511}^{\infty} N(E)dE \right] / \left[\int_{0}^{\infty} 
N(E)dE \right]$ (lower panel).  The horizontal dashed line indicates
the initial values for a 50 keV Maxwellian.  Again, we can see that a 
significant fraction of particles are accelerated into a nonthermal 
population for $\tau_c \leq 1$.

\section{Conclusions and Discussions}

We have studied particle acceleration in galactic black hole
accretion disk coronae via interactions between electrons and fast
mode waves---specifically, via the transit-time damping process.
Including Coulomb collisions, inverse Compton scattering and 
synchrotron losses, we show that particles with speeds higher than
the Alfv\'en speed can be accelerated out of the thermal background,
and we obtain steady state particle distributions composed of a
Maxwellian plus a nonthermal high energy tail extending into several
tens of MeV.  Detailed radiation modeling will be presented in a
forthcoming work and we expect that the Maxwellian and the nonthermal
tail of the particle distribution will be responsible for the
power-law spectra in the tens of keV and the high energy gamma-rays
observed from several GBHCs such as Cyg X-1 and GRO J0422,
respectively.

The generation of plasma wave turbulence in accretion disk
environments is a fairly unexplored topic, but it is reasonable to
suppose that the fast mode waves will be excited since it is an
intrinsic long-wavelength mode of a magnetized plasma. We
emphasize that the coronal plasma $\beta$ must be $< 1$ for electrons
to get most of the wave energy, otherwise proton acceleration becomes
possible, reducing the energy flow to the electrons.

The particle acceleration mechanism discussed here also has direct
implications on the high energy radiation from accretion disks in
AGNs, notably Seyfert galaxies. Preliminary analyses have indicated
that most of our results are insensitive to the size of the system;
thus, we expect $>$ MeV emissions are also being produced in Seyferts
as well, though high quality spectra above 200 keV are clearly needed
to firmly settle this issue.

\acknowledgments
We thank the referee for many constructive suggestions.
Comments from Drs. Edison Liang and Chuck Dermer are appreciated.
H.L. gratefully acknowledges the support of the Director's Postdoc 
Fellowship at LANL. Part of the work was completed during J.A.M.'s visit
at LANL, also supported by the same fellowship.

\clearpage
\begin{figure}
\caption{The time evolution of the electron density distribution as
a function of kinetic energy $E$ (upper panels) and wave spectral density
as a function of wavenumber $k$ (lower panels). Three coronal plasma
densities are considered. There are 16 curves in
each plot and the numbers indicate their time-sequence, spanning from
$t=0 - 0.6 \tau_{dyn}$, where $\tau_{dyn} \sim 1.5\times 10^{-3}$ sec.
The initial 50 keV Maxwellians are shown as the
thick-dashed curves (wave spectrum is zero initially) and the thick-solid
curves show the steady state particle and wave distributions.
As wave cascade progresses, the mean wavenumber becomes large enough 
that efficient acceleration results, as indicated by the development 
of the hard tails beyond the Maxwellian in particle distributions.
}
\end{figure}

\begin{figure}
\caption{The fraction of electrons with $E \geq 511$ keV out of
the total electron population at steady state for the coronal optical depth
$\tau_c$ ranging from $0.1$ to $1$, in both the energy content $f_{\rm E511}$
(upper panel) and the number density $f_{\rm N511}$ (lower panel).
The dashed lines indicate the values for a 50 keV Maxwellian.
When $\tau_c$ is high, electrons are mostly nonrelativistic due to
cooling, but a good fraction of electrons becomes relativistic 
and nonthermal from the wave acceleration when $\tau_c \leq 0.5$.
}
\end{figure}

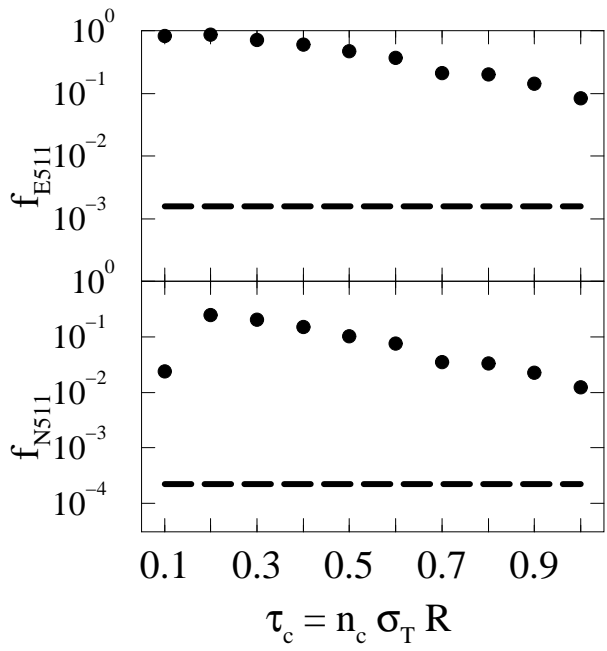
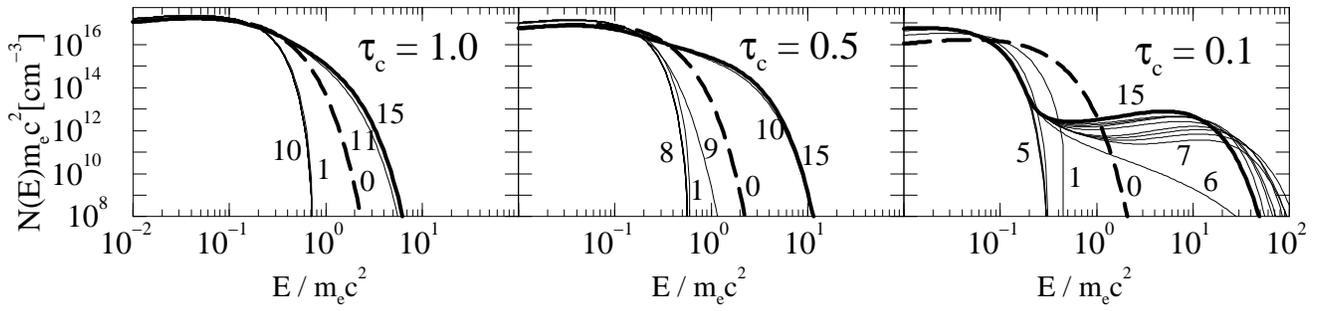
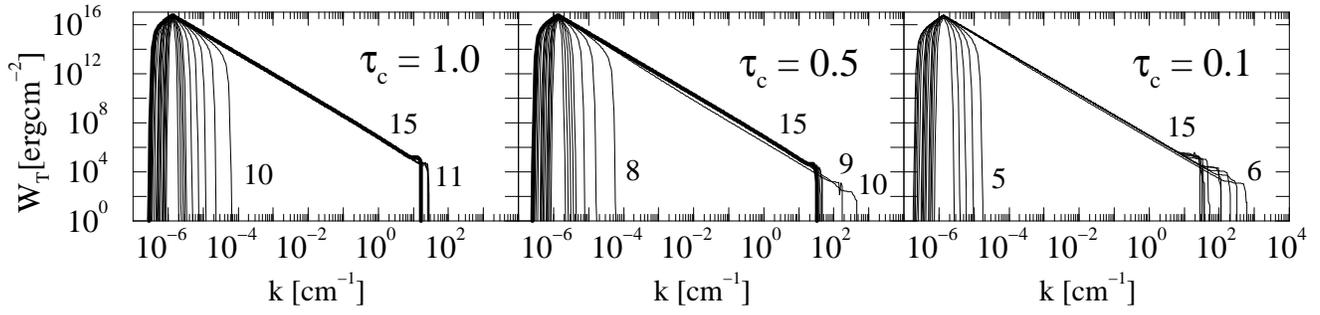
\end{document}